\documentclass[a4paper]{a2}
\usepackage{psfig,times}
\psfigurepath{/z/rue/plots/bonannoplot}

\begin{document}
\def\ni{\noindent} 
\def\ea{{et\thinspace al.}}                        
\def\eg{{e.g.}\ }                                    
\def\ie{{i.e.}\ }                                   
\def\cf{{cf.}\ } 
\def\''{\lq\lq}
\def\rot{\mathop{\rm rot}\nolimits}
\def\div{\mathop{\rm div}\nolimits}
\def\solar{\ifmode_{\mathord\odot}\else$_{\mathord\odot}$\fi} 
\def\gsim{\lower.4ex\hbox{$\;\buildrel >\over{\scriptstyle\sim}\;$}} 
\def\lsim{\lower.4ex\hbox{$\;\buildrel <\over{\scriptstyle\sim}\;$}} 
\def\~  {$\sim$} 
\def\cl {\centerline} 
\def\rl {\rightline} 
\def\x	{\times} 
\def\newpage{\vfill\eject} 
\def\bl{\par\vskip 12pt\noindent} 
\def\bll{\par\vskip 24pt\noindent} 
\def\blll{\par\vskip 36pt\noindent}
\def\alf{$\alpha$}
\def\L{$\Lambda$}
\def\Om{{\it \Omega}}
\def\nT{$\nu_T$\ }
\def\mT{$\mu_T$\ }
\def\cT{$\chi_T$\ }
\def\eT{$\eta_{\rm T}$}
\def\apj{{ApJ}}       
\def\apjs{{ Ap. J. Suppl.}} 
\def\apjl{{ Ap. J. Letters}} 
\def\mn{{MNRAS}} 
\def\aa{{A\&A}} 
\def\aas{{A\&AS}}
\def\ara{{ARA\&A}}
\def\aasup{{ Astr. Ap. Suppl.}} 
\def\csss{{Cool Stars, Stellar Systems, and the Sun} }
\def\an{{Astron. Nachr.}}
\def\sp{{Sol. Phys.}}   
\def\gafd{{GAFD}} 
\def\ass{{Ap\&SS}}
\def\acta{{Acta Astron.}}
\def\jfm{{J. Fluid Mech.}}
\def\nat{{Nat}}

\def\R{R\"udiger}
\def\E{Elstner}
\def\K{Kitchatinov}
\def\B{Brandenburg}
\def\F{Ferri\`{e}re}
\def\S{St\c{e}pie\`{n}}

\def\abs{\par\bigskip\noindent}            
\def\qq{\qquad\qquad}                      
\def\qqq{\qquad\qquad\qquad}               
\def\q{\qquad}
\def\DR{differential rotation\ }
\def\bib{\item{}}

\def\top{\item}
\def\toptop{\itemitem}
\def\so{$\Longrightarrow$\ }
\def\start{\begin{itemize}}
\def\stop{\end{itemize}}
\def\beg{\begin{equation}}
\def\ende{\end{equation}}

\title{Parity properties of an advection-dominated solar $\alpha^2\Om$-dy\-na\-mo}   
\author{A. Bonanno$^1$ \and D. Elstner$^2$ \and G. \R$^{2}$ \and G. Belvedere$^3$}
\offprints{alfio@ct.astro.it}
\institute{$^1$ Osservatorio Astrofisico di Catania, Via S.Sofia
78, I-95123 Catania, Italy\\
$^2$ Astrophysikalisches Institut Potsdam,  An der Sternwarte 16, 
D-14482 Potsdam, Germany\\
$^3$Dipartimento di Fisica ed Astronomia, Via S.Sofia 78, I-95123, Catania, Italy}
\date{\today}

\abstract{We have developed a high-precision code which solves the 
kinematic dynamo problem both for given rotation law and meridional flow in the case of 
a low eddy diffusivity of the order of $10^{11}$ cm$^2$/s known from the sunspot
decay. All our models work with an \alf-effect which is positive (negative) in the northern
(southern) hemisphere. It is concentrated in radial layers located either at the
top or at the bottom of the convection zone.
We have also considered an \alf-effect uniformly distributed in all the convection zone.
In the present paper the main attention is focused on i) the parity of the solution, ii)
the form of the butterfly diagram and iii) the phase relation of the resulting
field components. If the helioseismologically derived internal solar rotation law
is considered, a model without meridional flow of high magnetic Reynolds number (corresponding to
low eddy diffusivity) fails in all the three issues in comparison with the
observations. However, a meridional flow with equatorial drift at the bottom of the
convection zone of few meters by second can indeed enforce the equatorward migration of
the toroidal magnetic field belts similar to the observed butterfly 
diagram but, the solution has only a dipolar parity if
the (positive) \alf-effect is located at the base of the convection zone rather
than at the top. We can, therefore, confirm the main 
results of a similar study by Dikpati \& Gilman (2001).
}
\authorrunning{A. Bonanno et al.}
\titlerunning{Parity properties of  the 
circulation-dominated $\alpha^2\Om$-dy\-na\-mo}
\maketitle
\section{Introduction}
\begin{figure}
\psfig{figure=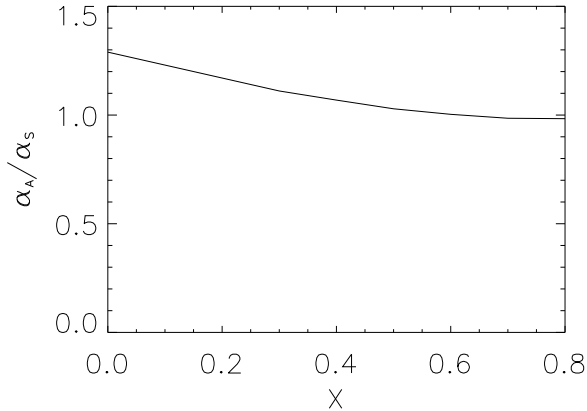,width=9cm,height=6.5cm}
\caption{The ratio of the dynamo numbers for equatorially antisymmetric 
(dipoles) to equatorially symmetric (quadrupoles) solutions for the shell dynamo 
of Roberts (1972). 
At $x_{\rm i}$ = 0.65 both values are equal. For thinner 
shells the most efficient dynamo is of dipolar type while for
ticker shells is of quadrupolar type.}
\label{ff1}
\end{figure}

In an early paper Roberts (1972) discussed the excitation conditions of
distributed shell dynamos. The \alf-effect, antisymmetric with respect to the
equator, was considered in an outer shell ($x_{\rm i} \leq x \leq 1$,  
$x=r/R_\odot$), the dynamo was
embedded in vacuum and the differential rotation $d\Om/dx$ was considered as
uniform throughout the outer shell. He found for positive dynamo numbers
($\alpha_{\rm north} \cdot d\Om/dx > 0$) that dipoles are more easily excited
than quadrupoles for thin convection zones whilst for deeper zones the quadrupoles
are more easily excited (see Fig.~\ref{ff1}).

These models are rather rough. The differences between the dipolar
solutions and the quadrupolar solutions always happen to be small, 
so that the important question concerning the latitudinal symmetry, 
that is the parity, did not seem to be a serious problem.

Roberts \& Stix (1972) computed $\alpha^2\Om$-dynamos with a rotation law
similar to the solar rotation law known nowadays by helioseismology. The results are
given in their Fig. 4b. Without latitudinal differential rotation ($\beta=0$)
but with $\partial \Om /\partial x\big|_{\rm eq} > 0$ the solution with quadrupolar
(symmetric) parity has a lower eigenvalue than the solution with the dipolar
symmetry (opposite to the Steenbeck-Krause (1969) model with $\partial
\Om/\partial x \big|_{\rm eq}<0$, see Fig. 3 of Roberts \& Stix). The
inclusion of the latitudinal shear produces still higher dynamo numbers but
{the difference between quadrupolar parity and dipolar parity grows}. 

K\"ohler (1973) thus considered only the excitation of modes 
with prescribed dipolar parity. For positive \alf-effect in 
the northern hemisphere and with an outwards increasing $\Om$,  only
a poleward  drift of  the toroidal magnetic field belts was found, contrary to the 
observations.

Moss \& Brooke (2000), in order to produce the observed equatorward migration of
the toroidal fields adopt the solar rotation law and a negative northern
\alf-effect in the bulk of the convection zone. The dipolar solutions are only
slightly easier to excite than the quadrupolar ones (dipole: $C_\alpha = -
3.20$, quadrupole $C_\alpha = - 3.25$). The parity problem does not seem to
exist for negative dynamo numbers (see also Fig. 3 in Roberts \& Stix 1972). The
situation, however, completely changes if a positive northern \alf-effect is
considered (Roberts \& Stix 1972, Fig. 4; Moss 1999). In this case one 
has to deal with a parity problem for the solar dynamo.

Following Dikpati \& Gilman (2001) we shall here study  the  parity problem for solar dynamos in particular 
for dynamos with rather small eddy diffusivities so that the meridional flow plays an important role in advecting the
toroidal magnetic field belts (Wang et al. 1991; Choudhuri et al. 1995). As the rotation law can be considered 
as given (known from helioseismology), we are free to vary the location of the \alf-effect, so 
that models are assumed to have a positive \alf-effect both at the top and at the bottom of the convection 
zone, as well as in the full convection zone. 

The inclusion of the meridional flow $u_{\rm m}$ 
has a strong impact on the mean-field dynamo when the eddy diffusivity $\eta_{\rm T}$
is low. In particular, for \eT$=10^{11}{\rm cm}^2$/s , as is known 
from the sunspot decay, the magnetic Reynolds number ${\rm Rm} = u_{\rm m} R_\odot /\eta_{\rm T}$
reaches values of the order of $10^3$ for a flow of $10$ m/s. As a consequence,
depending on the localization of the dynamo wave, a dramatic
modification of both magnetic field configurations and cycle period is expected. 
This possibility has recently been a subject of intense 
numerical investigation  (Dikpati \& Charbonneau 1999;
K\"uker et al. 2001), where it has been shown that solutions with high magnetic Reynolds number
provide correct cycle period, butterfly diagrams, magnetic phase relations and 
sign of current helicity, by means of 
a {\it positive} $\alpha$-effect in the north hemisphere.

In this investigation we study how the presence of the flow and the location of the
turbulent layer affect the parity mode selection and the cycle period. 
In this respect we show that, for high magnetic Reynolds numbers of the flow,
quadrupolar field configurations are more easily excited 
than the dipolar ones if there is no \alf-effect below $r/R_\odot \approx 0.8$. 
\section{Basic equations}
In the following the dynamo equations are given with the
inclusion of the induction by meridional circulation. 
Axisymmetry  implies that the mean flow in spherical coordinates is given by 
\beg
\vec{u}  = \left( u_r, u_\theta, r \sin\theta\ \Om\right).
\label{2.2}
\ende
In our formalism the magnetic field reads
\begin{equation}
\vec{B}  =  \left( {1\over r\sin\theta}{\partial A  \sin\theta
	     \over\partial\theta},
-{1\over r }{\partial A r\over\partial r}\ 
     ,\ B \right) ,
\label{2.3}
\end{equation}
where $A$ is the poloidal-field potential and $B$ is the toroidal field.
The dynamo equation may be written in the  form
\beg\label{eqd}
{\partial {\vec{B}} \over \partial t} = \rot ({\vec{u}}
\times {\vec{B}} + \alpha \vec{B}) - \rot \left(\sqrt{\eta_{\rm T}}\left(\rot
\sqrt{\eta_{\rm T}} {\vec{B}}\right)\right),
\label{2.70}
\ende 
which includes the  diamagnetism due to non-uniform turbulence. 
{ In fact it has been shown 
(\K \& \R, 1992) that the diffusive part of the 
mean turbulent electromotive force reads 
$ -\eta_{\rm T} 
{\rot} \vec{B} -\nabla\eta_{\rm T}\times\vec{B}/2 = 
-\sqrt{\eta_{\rm T}}{\rot}\sqrt{\eta_{\rm T}}\vec{B}$.
The second term is due to the turbulent diamagnetism.}
If there are strong gradients of turbulence intensity, 
this term will dominate the transport of the mean magnetic fields. 

As usual, the meridional circulation is derived from a stream function, 
so that the condition $\div \rho\vec{u} = 0$ is automatically fulfilled. 
Then a series expansion in Legendre polynomials is introduced, as described  in \R\ (1989): 
\begin{eqnarray}
\hat u_r&=& {1\over r^2\hat\rho \sin\theta} \ {\partial \psi \over
\partial \theta} = \sum\limits_{2,4} {1\over \hat \rho r^2} \ n(n+1)
\Psi_n \ P_n ,
\label{9}\\
\hat u_\theta&=& - {1\over r \hat \rho \sin\theta} \ {\partial \psi \over
\partial r} = - \sum\limits_{2,4} {1\over \hat \rho r} \ {d\Psi_n
\over dr} P_n^{(1)} .
\label{10}
\end{eqnarray}
A one-cell meridional circulation is described by
\begin{eqnarray}
u_r&=& {3\cos^2\theta -1 \over \hat\rho x^2} \psi,
\label{12}\\
u_\theta&=& - {\cos\theta \sin\theta \over \hat \rho x} {d\psi\over dx},
\label{13}
\end{eqnarray}
where $\psi$ is the usual stream function. A positive $\psi$ describes a cell 
circulating clockwise in the northern hemisphere, \ie the flow is 
polewards at the bottom of the convection zone and equatorwards at the surface.
For a negative $\psi$ the flow is, as is observed, polewards at the surface. In 
order to keep the flow inside the convection zone, the function $\psi$ must be zero at 
the surface and at the bottom of the convection zone. In order to define the strength of the
flow, we shall use the values $u_{\rm m}$ of the meridional circulation at the bottom
of the convection zone. 
\begin{figure}
\psfig{figure=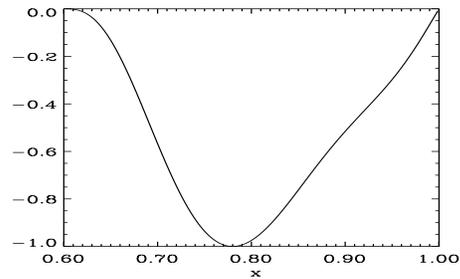,width=7cm,height=4cm}
\caption{The stream function $\psi$ of the meridional circulation used in the 
model computations. It must be negative if the circulation shall proceed 
equatorwards at the surface.} 
\label{f2}
\end{figure}
The rotation law for the solar dynamo can be considered as given 
by the helioseismic observations, in particular we have used the 
analytical model of Dikpati \& Charbonneau  (1999)
which is characterized by the existence of a steep subrotation
profile in the polar region with a thickness of about 0.05 solar
radii (see Markiel \& Thomas 1999). We then define the profile of \eT\ as
\beg\label{eta}
\eta_{\rm T} = \eta_c+{1\over 2}(\eta_t-\eta_c)(1+{\rm erf}(40(x-0.7))),
\ende
where $x=r/R_\odot$ is the fractional radius, {\em erf} denotes the error function,
$\eta_t$ is the eddy diffusivity, and $\eta_c$ the magnetic diffusivity beneath the
convection zone. The factor 40 defines the thickness of the transition region 
to be $0.05 R_\odot$, and we have adopted both 
$\eta_c/\eta_t = 0.1$ and $\eta_c/\eta_t =0.02$.

There is an increasing evidence for an \alf-effect in the convection
zone rather than in the overshoot region. Observations of the turbulent current 
helicity by Seehafer (1990), Pevtsov \ea\ (1995), Abramenko \ea\ (1996) as well as  Bao \& Zhang
(1998)  always lead to {\em negative} values in 
the northern hemisphere and  positive values in the southern hemisphere. If the 
observations indeed concern the current helicity of the {\em fluctuations} 
then after R\"adler \& Seehafer (1990) the azimuthal \alf-effect should be {\em positive}.
Also Brandenburg (1999) with new numerical simulations found (highly noisy)  positive   
correlations between the turbulent EMF and the mean magnetic field, \ie positive 
\alf-effect 
for the northern hemisphere. Our SOCA theory of Lorentz force-driven turbulence 
in a stratified rotating plasma also leads to negative current helicity and 
positive \alf-effect for the northern hemisphere (\R\ \ea\ 2001). 
Lastly, with their high-detailed simulation, Ossendrijver  et al. (2001) 
show that indeed far beyond the limits of the SOCA 
theory the results are confirmed: in the northern part of the solar convection zone 
the  horizontal  \alf-effect proves to be positive at variance with 
the vertical \alf-effect and the current helicity. 

\begin{figure}
\psfig{figure=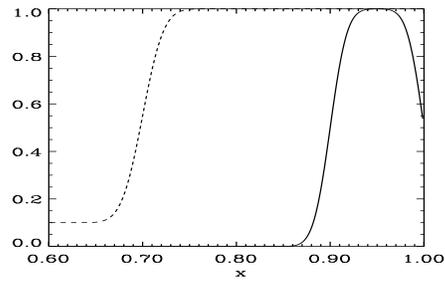,width=7cm,height=4cm}
\caption{The \alf-effect normalized function  in the bulk of 
the convection zone (solid line) and the profile of the eddy diffusivity 
(dashed line).} 
 \label{f1}
\end{figure}

The \alf-effect is always antisymmetric with respect to the equator, so that we 
write
\beg
\alpha = \alpha_0 \cos\theta\Big (1+{\rm erf}(\frac{x-x_\alpha}{d})\Big )\Big (1-{\rm erf}
(\frac{x-x_\beta}{d})\Big )/4,
\label{aa}
\ende
where $\alpha_0$ is the amplitude of the \alf-effect
and $x_\alpha$, $x_\beta$ and $d$ define the location and the thickness of the 
turbulent layer, respectively. Differently from the overshoot 
dynamo $\alpha_0$ is {\em not} assumed to change its sign in the bulk of the 
convection zone or in the overshoot layer.
In Fig. \ref{f1} the eddy diffusivity profile and the $\alpha$ profile are shown
for $x_\alpha=0.9$,  $x_\beta=1$ and $d=0.05$ which correspond to 
an \alf-effect located at the top of the convection zone. 
\begin{figure}
\psfig{figure=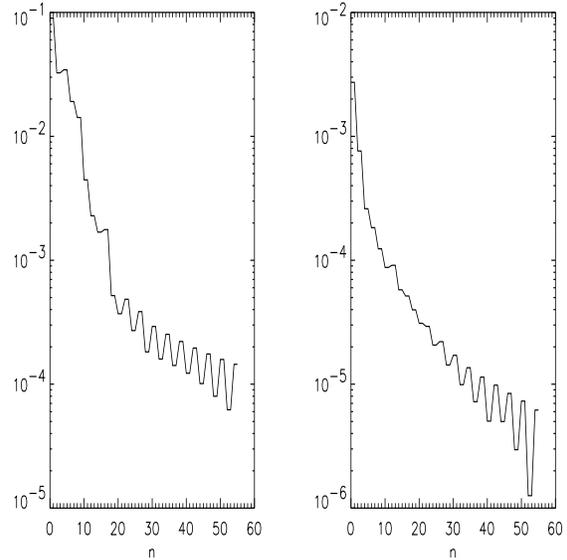,width=8.0cm,height=8.0cm}
\caption[]{Truncation dependence of a typical high-Reynolds number solution. 
The quantities $\max_x \{b_n\} $ and $\max_x \{a_n\}$ as a function of $n$ are shown,  
in the left and in the right panel, respectively.} 
\label{conv}
\end{figure}
\section{Mathematics and numerics}
In order to investigate the properties of the solar 
dynamo with such a low eddy diffusivity we
solve the linear dynamo Eq. (\ref{eqd}) with a finite-difference
scheme for the radial dependence and a polynomial expansion for the 
angular dependence.
In particular, we have used the following expansions for the field:
\begin{eqnarray}
\hat A(x,\theta)&= e^{\lambda t}& \sum\limits_{n} a_n(x)\ P_n^{(1)}(\cos\theta), 
\label{4.1}\\
\hat B(x,\theta)&=e^{\lambda t}& \sum\limits_{m} b_m(x)\ P_m^{(1)}(\cos\theta),
\label{4.2}
\end{eqnarray}
where $\lambda$ is the (complex) eigenvalue,
$n=1,3,5,\dots$ and $m=2,4,6,\dots$ for antisymmetric modes,
and $n\leftrightarrow m$ for symmetric modes.
Vacuum boundary conditions at the surface are then translated into 
\beg 
{d a_n\over dx} + (n +1)\;a_n = b_m = 0.
\label{4.3}
\ende
In the interior at $x=x_{\rm i}=0.6$ we have instead set 
\beg
x{d b_m \over d x} +b_m = a_n = 0,
\label{4.4}
\ende
which imply perfect conductor boundary conditions for the field.  

By substituting (\ref{4.1}) and (\ref{4.2}) in (\ref{eqd}) one obtains a set of infinite 
o.d.e. that can be conveniently truncated in $n$ when the desired accuracy is 
achieved. The system is in fact solved by means of a second order accuracy 
finite difference scheme and the basic computational task is thus
to numerically compute eigenvalues and eigenvectors of a 
block-band diagonal real matrix of dimension
$M\times n$, $M$ being the number of mesh points and $n$ the number of harmonics. 
This basic algorithm is embedded in a bisection procedure in order to 
determine the critical \alf-value needed to find a purely oscillatory solution. This 
value is accepted when the dimensionless quantity 
$\Re e(\lambda) R^2_\odot /\eta_{\rm T} $ is no greater than  
$10^{-3}$. 

We have first tested the accuracy of our code in simple cases where the 
eigenvalues and eigenfunctions are known (decay modes with $\alpha=0$ and 
simple analytic solutions with constant \alf). Then, we have checked most of 
the solutions discussed in the literature (no flow) and we found good
agreement. In fact we have rapid convergence in $n$ for 
simple (no-flow) $\alpha\Om$-dynamo, and we could confirm most of the 
findings by Steenbeck \& Krause (1969) and Roberts (1972).
When the flow is present, 
the number of harmonics needed in order to get convergence generally increases
with the Reynolds number and we decided to truncate the series 
(\ref{4.3}-\ref{4.4}) when the maximum value over $x$ of the $n$-th harmonic drops by 
roughly three orders of magnitude, as shown in Fig. \ref{conv}.
\section{Numerical results}
We performed an extensive numerical exploration of the 
parameter space, allowing simultaneous variations of 
$x_\alpha$, $\eta_c/\eta_t$, $u_{\rm m}$ and angular dependence
of the $\alpha$-effect, considering both $\cos\theta$ and 
$\cos\theta\sin^2\theta$ functions in (\ref{aa}). 
In all models discussed in the following 
the rotation law is given by helioseismology and the eddy 
diffusivity is fixed to be $\eta_{\rm T}=10^{11} \; {\rm cm}^2/{\rm s}$. 
The drift amplitude is considered as a free parameter 
of the order of 1 m/s up to 10 m/s. 
In some cases we have investigated the field configuration for very 
strong flow up to 20 m/s.
\subsection{Alpha-effect in the entire convection zone}
We start with a model where the (positive) \alf-effect exists through the whole
convection zone ($x_\alpha=0.7$, $x_\beta=1$, $d=0.05$). 
In Table \ref{tab1} the results are given. Suffix A denotes the
(dipolar) solutions with antisymmetry with respect to the equator and suffix S
denotes the (quadrupolar) solutions with symmetry with respect to the equator.
The drift amplitude varies between 2 cm/s and 10 cm/s. 
For small flow the dipole-solution occurs with the lowest \alf-effect amplitude, but for
higher values of the flow the quadrupole solution is more easily excited. 
The oscillation period of the linear dynamo is also given:
for high value of the flow  quadrupole solutions have
cycles which are longer than the corresponding dipole ones. 
This is a property that we 
have verified for all cases in the rest of paper, namely that in
the high Reynolds number regime, 
lowest \alf-value solutions have longer periods than the 
corresponding opposite parity solutions.

\begin{figure}[h]
\psfig{figure=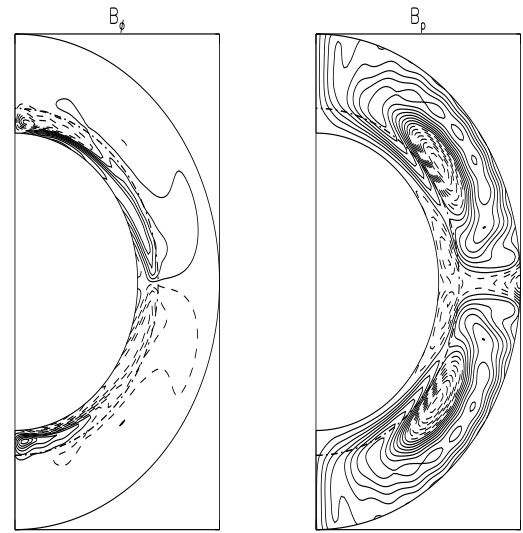,width=8.0cm,height=8.0cm}
\caption[]{Alpha-effect in the entire convection zone: Toroidal (left) and poloidal (right) 
antisymmetric (dipolar) field configuration for 
$u_{\rm m}=6$ m/s at $t=0$. 
The \alf-effect is located between $x_\alpha=0.7$ and $x_\beta=1$. 
The solution is plotted as $B_\phi$-isocontours and poloidal magnetic field lines.
{Solid contours (field lines)
correspond to positive toroidal fields (pointing into the plane
of the paper) and dashed contours to negative
toroidal field (pointing out of the paper)} .} 
\label{entire1}
\end{figure}

\begin{figure}[h]
\psfig{figure=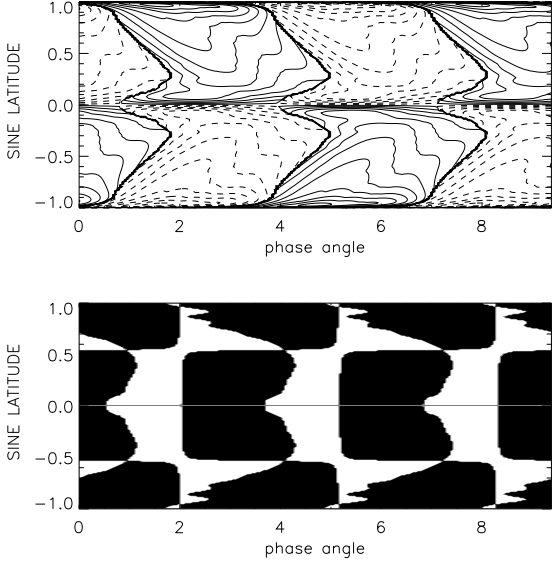,width=8.0cm,height=8.0cm}
\caption[]{Alpha-effect in the entire convection zone: Butterfly diagram (top) 
and field phase relation (bottom) of a dynamo with 
critical turbulence
$\alpha_0 = 2.46$ cm/s, 
$u_{\rm m}=6$ m/s.
Black colors means negative $B_r B_\phi$.
The \alf-effect is located between $x=0.7$ and $x=1$. } 
\label{entire2}
\end{figure}

It should also be noticed that, with the small value of
eddy diffusivity we have chosen, the cycle period becomes rather long compared
with the 22-years of the Sun. As we know from the theory of the overshoot
dynamo, the inclusion of the nonlinear feedback of the magnetic field (via
\alf-quenching) reduces the periods to more realistic values (R\"udiger
\& Brandenburg 1995). What we are interested in, is mainly the 
influence of the \alf-effect profile and the drift amplitude on the resulting
parity of the solutions. 
Figures \ref{entire1} and \ref{entire2} show the magnetic topology of the dynamo
with 3 m/s drift amplitude. We find (for the lowest eigenvalue) a solution
with antisymmetry with respect to the equator. The toroidal field belts are
concentrated at the bottom of the convection zone, and the poloidal field
exhibits, close to the surface, a rather small-scaled structure (Fig. \ref{entire1}). 
The toroidal field belts migrates equatorwards but, however, the maximal field
strength occurs in the polar region. 
\begin{table}[h]
\caption[]{Critical \alf-values (cm/s) and periods $P$ (yrs) for models with
$\alpha$-effect in the entire convection zone. 
Bold is used when the dipolar solution has the smallest  \alf-value. 
The quadrupolar symmetry is denoted
by ``S'' (``symmetric'') and the dipolar parity symmetry by ``A'' (antisymmetric). }
\begin{tabular} [t]{|ccccc|}
\hline
 $u_{\rm m} $ & $\alpha_{\rm A}$ &  $P_{\rm A}$     &  $\alpha_{\rm S}$ &
 $P_{\rm S}$   \\
\hline
$2.0$  &  {\bf 0.90}      &  $\infty$  &    1.28     & 131     \\
$3.0$  &  {1.83}      &  82  &    {1.70}     & 83     \\
$6.0$  &  {2.46}      &  51  &    {2.17}   & 54 \\
$10.0$  &  { 3.93}      &  37       &    {3.45}     & 41   \\
\hline
\end{tabular}
\label{tab1}
\end{table}
\subsection{Alpha-effect at the top}
\begin{figure}
\psfig{figure=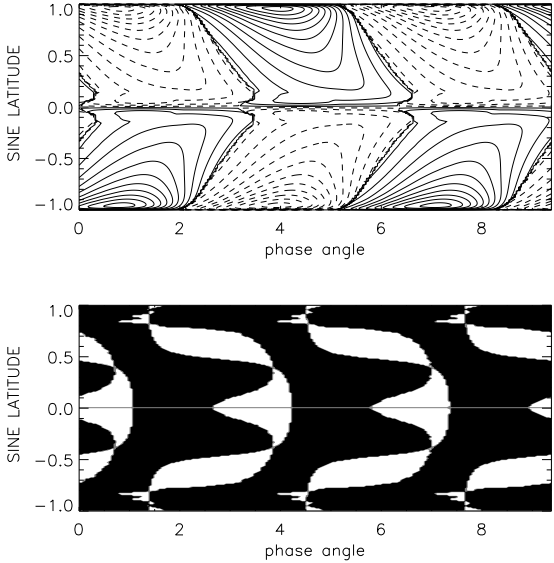,width=8.0cm,height=8.0cm}
\caption[]{Alpha-effect at the top: Butterfly diagram (top) and field phase relation (bottom) of a dynamo with 
critical turbulence
$\alpha_0 = 10.8$ cm/s, $u_{\rm m}=10$ m/s, $x_\alpha=0.9$, $x_\beta=1$.
Black colors means negative $B_r B_\phi$. }
\label{but}
\end{figure}

\begin{figure}
\psfig{figure=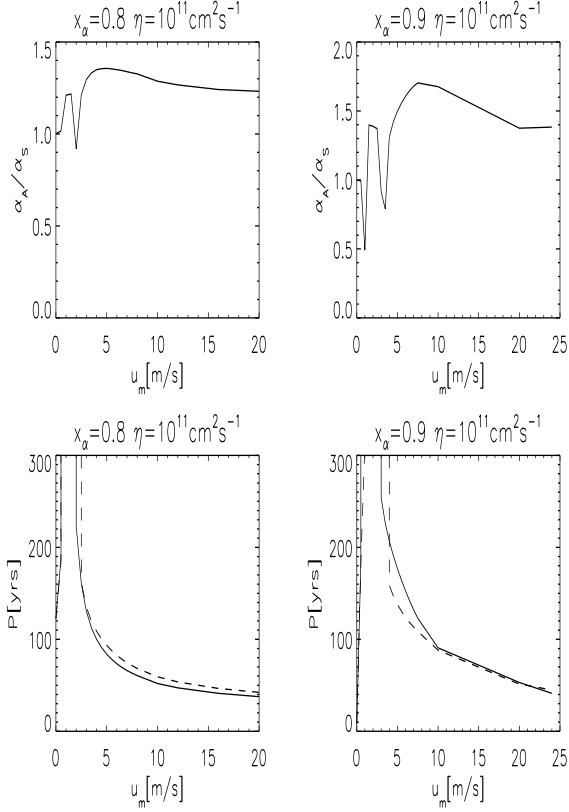,width=8.0cm,height=11.0cm}
\caption[]{Alpha-effect at the top: The ratio $\alpha_A/\alpha_S$ of the critical 
\alf-numbers for antisymmetric and symmetric field configurations 
is shown (as in Fig. \ref{ff1}) as a function of the flow speed $u_{\rm m}$ in the panels above. 
The period (continuous line for the antisymmetric modes, dashed line for symmetric modes)
is shown in panels below. The presence of vertical lines indicates that the 
solution is stationary }
\label{rate}
\end{figure}

We now consider models where the \alf-effect is located at the top of the convection 
zone as discussed in the models of K\"uker et al. (2001). 
The results of the simulations confirm the basic features 
of the advection-dominated dynamo, namely that,   
for a flow of few m/s and a low diffusivity, the butterfly diagram shows the correct equatorward migration 
of the toroidal field, and the phase relation of the magnetic fields  is 
mostly negative as shown in Fig. \ref{but}. 
We found that, as far as the parity model selection is concerned, 
variations of the ratio $\eta_c/\eta_t$ are not particularly
significant, while a functional dependence of the type $\cos\theta\sin^2\theta$
for the \alf-effect disfavours the appearance of dipolar
field configurations. We have then considered variations of the thickness
of the $\alpha$-layer. The results of this investigation are summarized in 
Fig. \ref{rate} and in Tables \ref{tab2} and \ref{tab3}. 
\begin{table}[h]
\caption[]{Alpha-effect at the top (thick layer): Critical \alf-values (cm/s) and periods (yrs) for various values 
of the flow (m/s), for both antisymmetric (A) and symmetric field (S) 
configurations with  $x_\alpha = 0.8$, $x_\beta=1$,
$\eta_c/\eta_t = 0.02 .$}
\begin{tabular}[t]{|ccccc|}
\hline
 $u_{\rm m} $ & $\alpha_{\rm A}$ &  $P_{\rm A}$     &  $\alpha_{\rm S}$ &
 $P_{\rm S}$   \\
\hline
$1.0$  &  0.96      &  $\infty$  &    0.79     & $\infty$     \\
$1.5$  &  1.56      &  $\infty$  &    1.28     & $\infty$ \\
$ 2.0$  &  {\bf 1.86}      &  222       &     2.02     & $\infty$ \\
$2.5$  &  2.22      &  160       &    1.83     & 162 \\
$3.0$  &  2.58      &  130       &    1.99     & 134 \\
$3.5$  &  2.95      &  112       &    2.21     & 119 \\
$5.0$  &  4.03      &  83        &    2.97     & 100   \\
$10 $  &  7.30      &  51        &    5.67     & 59 \\
$20$   &  12.1      &  37        &    9.86     & 42\\
\hline
\end{tabular}
\label{tab2}
\end{table}
\begin{table}[h]
\caption[]{Alpha-effect at the top (thin layer): Critical \alf-values (cm/s) and periods 
(yrs) for various values  of the flow (m/s), 
for both antisymmetric and symmetric field configurations 
with $x_\alpha = 0.9$, $x_\beta=1$, 
and $ {\mathbf \eta_c/\eta_t} = 0.02$ . The dipole solutions with 
lowest critical \alf-value are shown
in bold.}
\begin{tabular}[t]{|ccccc|}
\hline
 $u_{\rm m} $ & $\alpha_{\rm A}$ &  $P_{\rm A}$     &  $\alpha_{\rm S}$ &
 $P_{\rm S}$   \\
\hline
$1.0$  &  4.44      &  $\infty$  &    9.05     & 312     \\
$1.5$  &  5.63      &  $\infty$  &    4.03     & $\infty$ \\
$2.0$  &  7.11      &  $\infty$  &    5.11     & $\infty$ \\
$2.5$  &  8.79      &  $\infty$  &    6.41     & $\infty$ \\
$ 3.0$  &  {\bf 7.24}      &  253       &     7.93     & $\infty$ \\
$ 3.5$  &  {\bf 7.63}      &  226       &    9.61     & $\infty$ \\
$5.0$  &  9.88      &  176       &    6.59     & 137   \\
$10 $  &  19.1      &  90        &    11.4     & 88 \\
$24$   &  39.1      &  41      &    28.8     & 44\\
\hline
\end{tabular}
\label{tab3}
\end{table}
For a slow flow, both quadrupoles and dipoles have similar 
excitation conditions. For intermediate values of the flow amplitude,
it is possible to have a different dynamo mechanism working
and the solutions are stationary. 
In particular, as it is shown in Tables \ref{tab2} and \ref{tab3},
dipole solutions may have a smaller critical $\alpha_0$-value in this case.

\begin{figure}
\psfig{figure=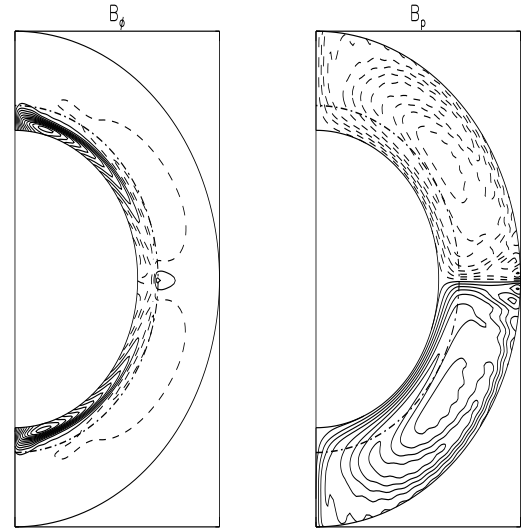,width=8.0cm,height=8.0cm}
\caption[]{Alpha-effect at the top: Toroidal (left) 
and poloidal (right) symmetric (quadrupolar) field  configuration for 
$\alpha_0 = 8.95$ cm/s, $x_\alpha=0.9$, $x_\beta=1$, $d=0.02$, ${\mathbf \eta_c/\eta_t} = 0.1$,
$u_{\rm m}=10$ m/s at $t=0$}
\label{f2000us}
\end{figure}

In the high-Reynolds numbers regime, quadrupole fields are more easily excited and the cycle 
periods drastically reduce.  
We confirm the findings by Dikpati \& Charbonneau (1999) as far as the global structure of the field and 
the parity problem (Dikpati \& Gilman 2001) is concerned.
For a flow of 10 m/s the field geometry is shown in Fig. \ref{f2000us} 
and the mode with the lowest eigenvalue is a quadrupole.
\subsection{Alpha-effect at the bottom}
\begin{figure}
\psfig{figure=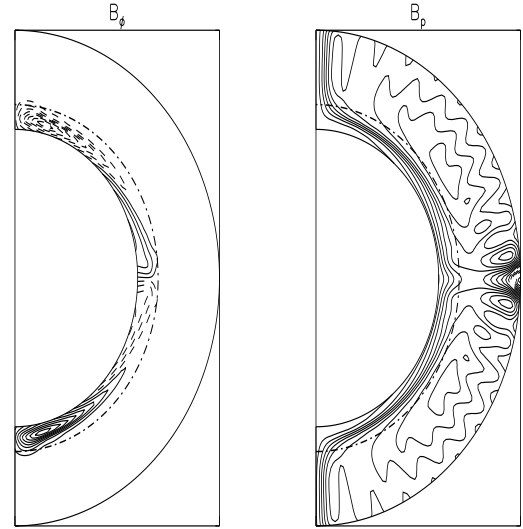,width=8.0cm,height=8.0cm}
\caption[]{Alpha-effect at the bottom: Toroidal (left) and poloidal (right) 
antisymmetric (dipolar) field configuration for \alf= 0.95 cm/s, 
$u_{\rm m}= 6$ m/s at $t=0$. 
The \alf-effect is located between $x_\alpha=0.7$ and $x_\beta=0.8$.} 
\label{f9}
\end{figure}
\begin{figure}
\psfig{figure=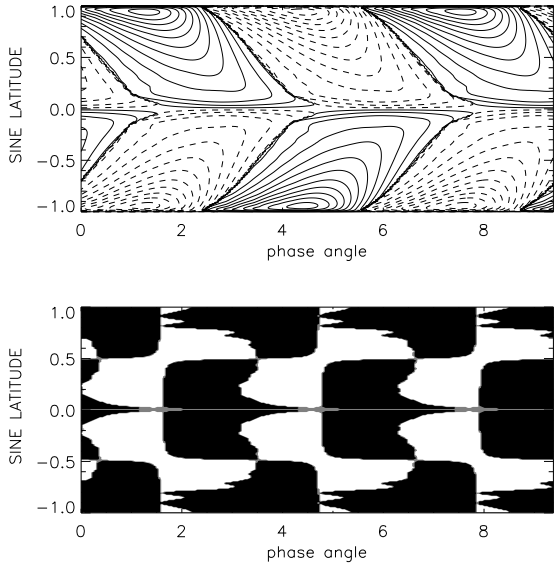,width=8.0cm,height=8.0cm}
\caption[]{The same as in Fig. \ref{but} 
but for the model presented in Fig. \ref{f9}. The cycle period is 120 years.} 
\label{f10}
\end{figure}
This picture drastically changes if the \alf-effect operates at the bottom
of the convection zone. We have considered a thin and a thick \alf-layer. In
the first class of models $x_\alpha=0.7$ and $x_\beta=0.8$
while in a second class $x_\alpha=0.7$ and $x_\beta=0.9$. 

Figures \ref{f9} and \ref{f10} show an example of a model of the first class.
Again, the toroidal field is concentrated at
the bottom of the convection zone, where, however, the highest field amplitudes
occur in the polar regions. The diagram for $B_r \cdot B_\phi$ shows the
dominance of the negative sign.

\begin{figure}
\psfig{figure=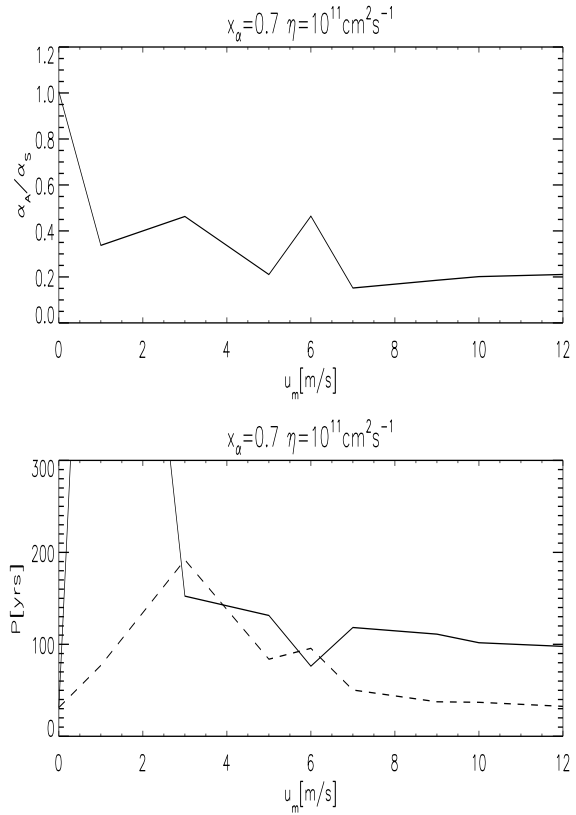,width=8.0cm,height=11.0cm}
\caption[]{The same as in Fig. \ref{rate} but for models with the \alf-effect 
located at the bottom of the convection with $x_\alpha=0.7$ and $x_\beta=0.8$. Note the clear dominance of 
dipolar solutions indicated by their much smaller dynamo numbers.}
\label{rate1}
\end{figure}

In Fig. \ref{rate1} parity and cycle periods of the solution for the model with a thin
\alf-layer at the bottom of the convection zone are given. The solution
with the dipolar symmetry has the lowest \alf-value and is 
the stable one. There is no heavy parity problem if the \alf-layer is located
at the base of the convection zone (see Dikpati \& Gilman 2001). However, we can
also notice from Fig. \ref{rate1} (lower panel) that in this case the oscillation period
of the dipoles is longer than for the quadrupoles, while, for slow flows, the
cyclic behaviour of the dynamo solution disappears, so that only meridional flows
with amplitudes exceeding 3 m/s are here relevant. In this case the dipole
solutions do not match the 22-year cycle period of the Sun (although the
quadrupoles do). The overall result is that the dipolar
solutions are always more easily excited and also the butterfly diagram shows
the correct characteristics.

In the second class of models the dynamo mechanism was of the same type, as
we could infer from the field configuration and from the butterfly diagram
(not shown).
However in this case the quadrupolar solution was always the most easily excited. 
From these results we can deduce that the region where the \alf-effect produces 
more easily dipolar field configuration, in the advection-dominated
regime,  is below $x\approx 0.8$. 
\subsection{Very thin \alf-layer}
\begin{figure}
\psfig{figure=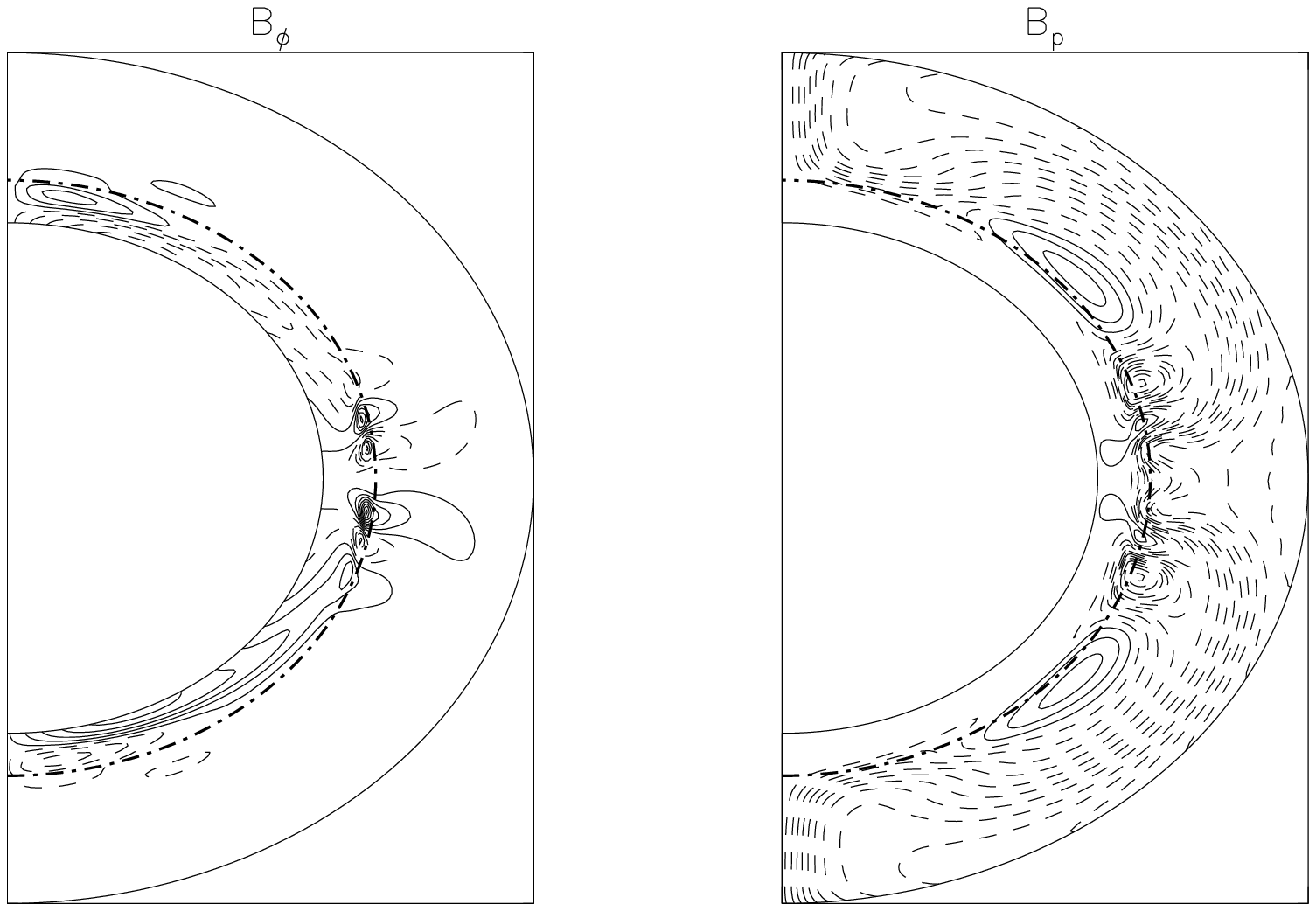,width=8.0cm,height=8.0cm}
\caption[]{Thin Alpha-effect at the bottom: Toroidal (left) and poloidal (right) 
antisymmetric (dipolar) field configuration for \alf= 4.36 cm/s, 
$u_{\rm m}=3$ m/s at $t=0$ and $x_\alpha=0.7$, 
$x_\beta=0.75$.} 
\label{thinf}
\end{figure}

\begin{figure}
\psfig{figure=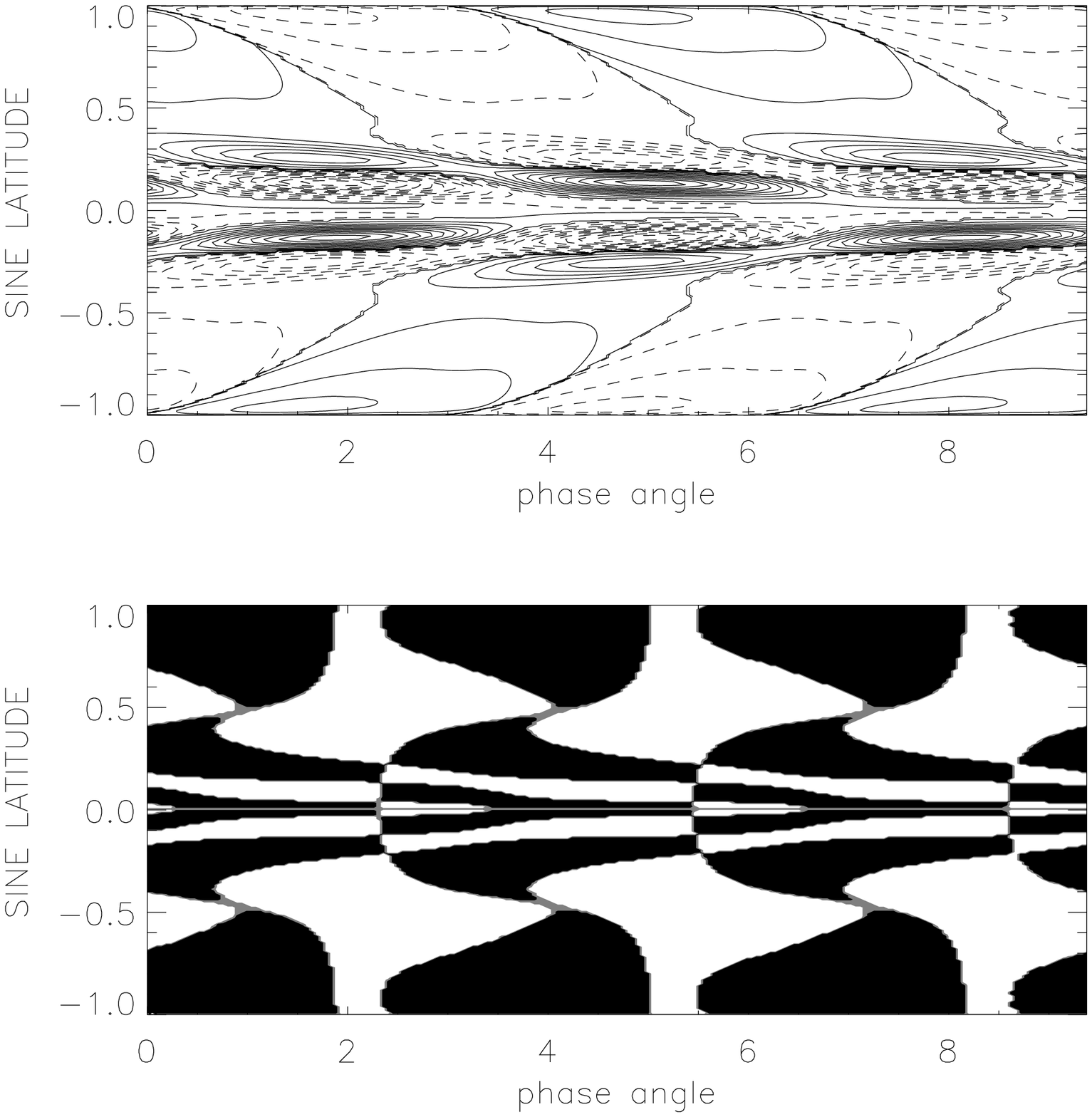,width=8.0cm,height=8.0cm}
\caption[]{The same as in Fig. \ref{but} but for the model presented in 
Fig. \ref{thinf}. The cycle period is 74 years.} 
\label{thinb}
\end{figure}

\begin{figure}
\psfig{figure=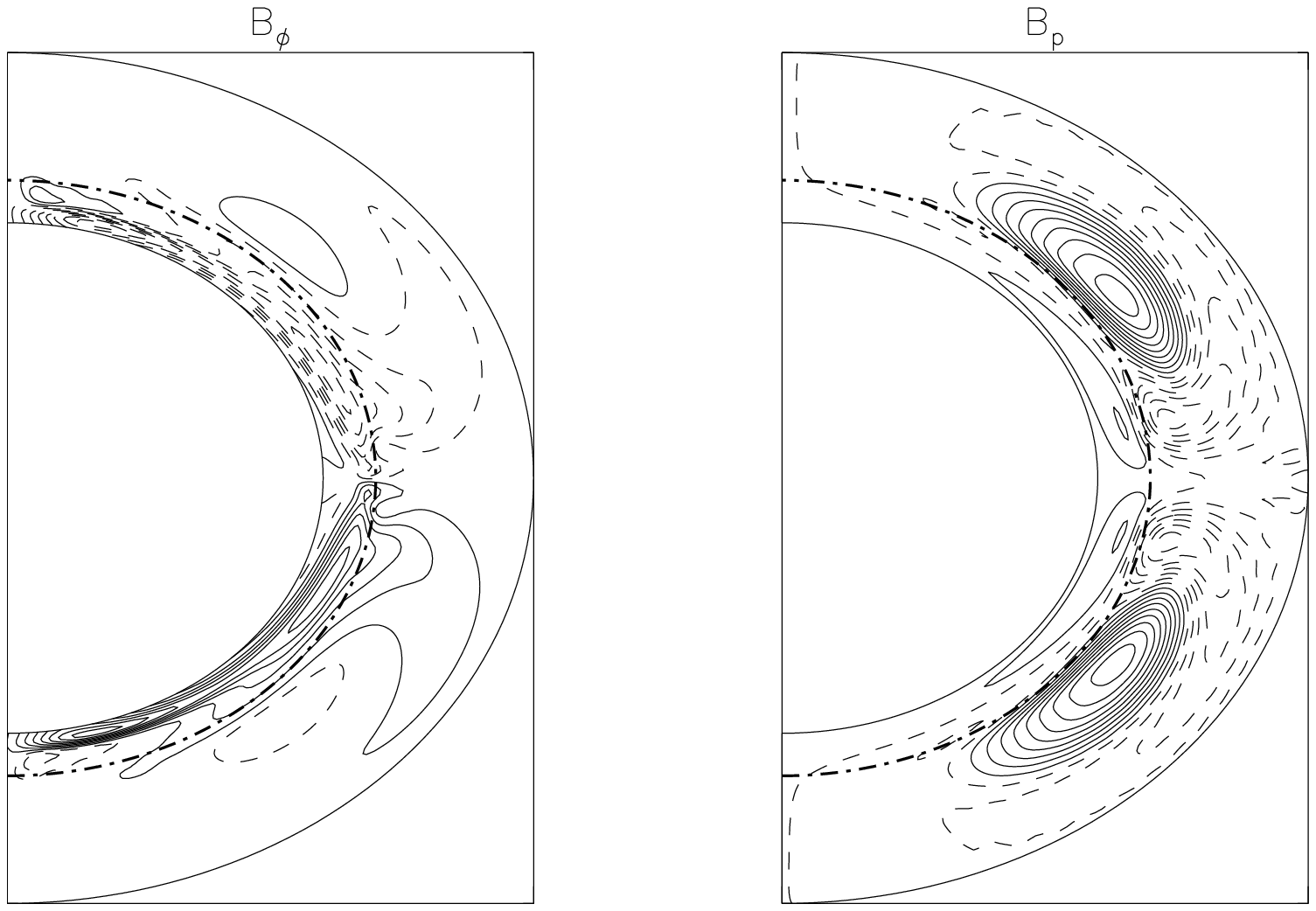,width=8.0cm,height=8.0cm}
\caption[]{Alpha-effect at the bottom for a very thin layer: toroidal (left) and poloidal (right) 
antisymmetric (dipolar) field configuration for \alf= 16.61 cm/s, 
$u_{\rm m}=12$ m/s at $t=0$ and $x_\alpha=0.7$, $x_\beta=0.75$ .} 
\label{bestf}
\end{figure}

\begin{figure}
\psfig{figure=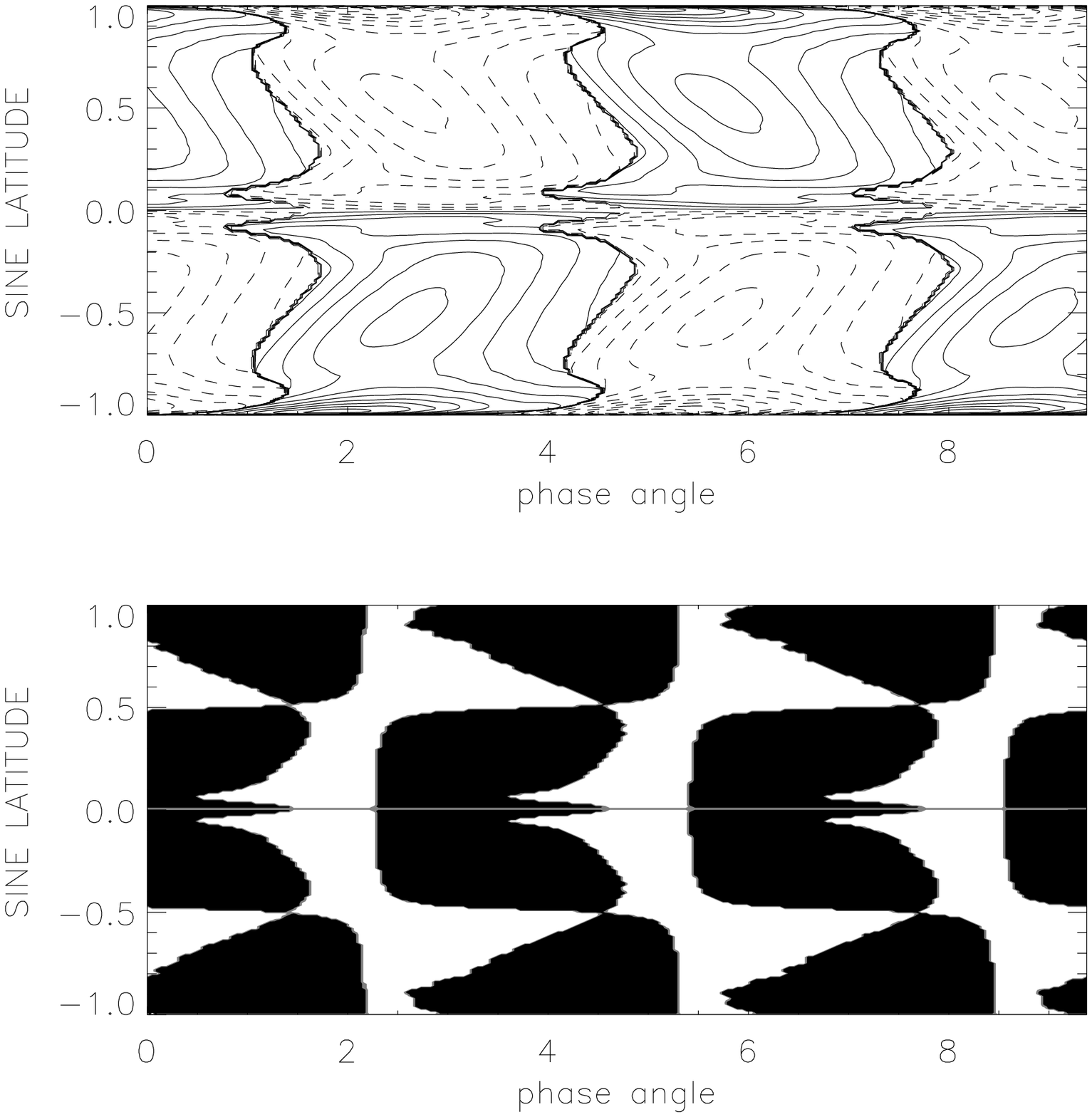,width=8.0cm,height=8.0cm}
\caption[]{The same as in Fig. 
\ref{but} but for the model presented in 
Fig. \ref{bestf}. The cycle period is 23 years.} 
\label{bestb}
\end{figure}

In this case we have set $x_\alpha=0.7$, $x_\beta=0.75$, $d=0.02$ and the relevant results are 
summarized in Table \ref{tab4}, 
where it is possible to notice that for a thin \alf-layer at the bottom, the dipolar
field configuration is the most easily excited. We noticed that if no flow is present
there is no migration towards lower latitudes of the magnetic field. However for moderate flow
the maximum of the toroidal field is in the equatorial region, as is possible to see in 
Fig.\ref{thinf} and Fig.\ref{thinb}. For stronger flow the period matches the observed one 
as shown in Table \ref{tab4}. In this case 
dipolar and quadrupolar solutions have greater dynamo numbers than in the 
slow flow case, but the dipolar field configuration is always
the favoured one. The field configuration and the butterfly diagram 
are shown in Fig. \ref{bestf} and \ref{bestb}.
Conversely, a very thin \alf-layer located at the top of the 
convection zone produces a much larger dynamo number and the solution is always symmetric 
as shown in Table \ref{tab4}.


\begin{table}[h]
\caption[]{Critical \alf-values (cm/s) and periods (yrs) for a very thin
$\alpha$-layer located between  $x_\alpha=0.7$ and $x_\beta=0.75$  for various values of the flow
(m/s), for both antisymmetric and symmetric field configurations. 
In the last line  the same $\alpha$-layer is located
between $x_\alpha=0.95$ and $x_\beta=1$ and the solution is clearly of the symmetric type. 
Also notice the much larger dynamo number in this latter case. }
\begin{tabular}[t]{|ccccc|}
\hline
 $u_{\rm m} $ & $\alpha_{\rm A}$ &  $P_{\rm A}$     &  $\alpha_{\rm S}$ &
 $P_{\rm S}$   \\
\hline
$1$  &  {\bf 0.43}      &  $\infty$  &    1.14     & 90.8     \\
$2$  &  {\bf 1.37}      &  $\infty$  &    1.83     & $\infty$ \\
$3$  &  {\bf 4.36}      &  74  &    4.94     & 64 \\
$5$  &  {\bf 7.53}      &  38      &    8.46     & 35   \\
$7$  &  {\bf 9.22}      &  34      &    9.65     & 33   \\
$12$  &  {\bf 16.61}      &  23      &    16.77     & 23   \\
\hline
$ 2$  &   87      &  $\infty$       &    {50}     & 283   \\
$ 20$  &   43      &  94       &    {40}     & 68   \\
\hline
\end{tabular}
\label{tab4}
\end{table}

\section{Conclusions}
A kinematic solar dynamo has been studied with the flow pattern taken from observations
(rotation law) and theory (meridional flow). An eddy diffusivity of $10^{11}$
cm$^2$/s provides us with a value consistent with the sunspot decay. With such a small
value the magnetic Reynolds number for a meridional flow of (say) 10 m/s reaches
values of order of 10$^3$, so that the dynamo can really be called 
advection-dominated. The meridional flow at
the bottom of the convection zone is supposed to drift equatorwards (and polewards at its top)
as can be deduced by the mean-field theory of the differential rotation
in an outer convection zone (Kitchatinov \& R\"udiger 1999; Miesch et al. 2000).
When this happens, the meridional flow advects the field equatorwards producing a
butterfly diagram of the observed type, which would not occur with i) a positive
\alf-effect (in the northern emisphere), ii) the standard rotation law known 
from helioseismology and iii) no meridional flow (Choudhuri et al. 1995; Dikpati \& Charbonneau 1999; K\"uker et al.
2001). As Dikpati \& Gilman (2001) have stressed, these models 
may encounter a problem with the parity of the solution 
since for most of the models the quadrupolar solution is the stable one. We have
used this striking fact to discuss the effect of the radial distribution of the \alf-effect 
on the parity model selection. Our conclusion is that a thin \alf-layer located below 
$r/R_\odot \approx 0.8$ selects models with correct parity property, butterfly diagram 
phase relation and cycle periods close to the observed one. 
This confirms the findings of Dikpati \& Gilman (2001), providing further evidence for 
a tachocline \alf-effect as a promising candidate for understanding the dynamo mechanism 
operating in the Sun. 

\end{document}